\documentstyle[aps,prl,twocolumn,graphics,epsfig,floats]{revtex}

\newcommand{\be}{\begin{eqnarray}}
\newcommand{\ee}{\end{eqnarray}}

\begin{document}

\title{
Polarization Entanglement Purification using Spatial Entanglement }

\author{Christoph Simon$^{1}$ and Jian-Wei Pan$^{2}$}
\address {$^1$
Centre for Quantum Computation, University of Oxford, Parks Road, Oxford OX1 3PU, United Kingdom\\  $^2$ Institut
f\"ur Experimentalphysik, Universit\"at Wien, Boltzmanngasse 5, A-1090 Wien,  Austria }

\date{\today}
\maketitle

\begin{abstract} Parametric down-conversion can produce photons that are entangled both in polarization and in
space. Here we show how the spatial entanglement can be used to
purify the polarization entanglement using only linear optical
elements. Spatial entanglement as an additional resource leads to
a substantial improvement in entanglement output compared to a
previous scheme. Interestingly, in the present context the thermal
character of down-conversion sources can be turned into an
advantage. Our scheme is realizable with current technology.

\pacs{03.67.Lx, 03.65.Bz}
\end{abstract}

Entanglement is an essential resource for quantum communication. It inevitably becomes degraded when the entangled
particles propagate away from each other. Entanglement purification \cite{bennett} is therefore essential for the
implementation of quantum communication over all but very modest distances . Entanglement purification describes
methods to generate close to maximally entangled pairs out of a larger number of less perfectly entangled pairs
using only local operations and classical communication. Long-distance quantum communication protocols such as
quantum repeaters \cite{briegel} require many purification steps to establish entangled pairs of good quality
between distant locations. If the local operations are sufficiently precise, then secure quantum communication is
possible over arbitrary distances \cite{aschauer}. Therefore simple and precise implementations of entanglement
purification are very desirable.

Photons are the best physical systems for the long-distance transmission of quantum states. Purification methods
for entangled photons are therefore of particular interest. We have recently proposed such a scheme, which uses
only linear optical elements \cite{pan}. In \cite{knill} a way of realizing quantum computation with linear optics
was suggested. These schemes were designed for ideal photon or photon-pair sources, i.e. for sources that produce
at most one photon or one pair of photons at a given time.

However, at the moment parametric down-conversion (PDC) \cite{pdc} is still the best source of entangled photons.
PDC is not an ideal pair source, but it is quasi-thermal: if the probability to emit one pair of photons at a given
instance is of order $p$, then the probability to emit two pairs is of order $p^2$ etc. This is usually perceived
as a problem, because it means that some quantum information protocols that would work for single-pair sources fail
for PDC sources \cite{kok}. Here we will show that not only is entanglement purification with linear optics still
possible for PDC sources, but their characteristics can even be turned into an advantage. The main reason for this
is that in PDC both polarization and spatial entanglement can be produced naturally, and the spatial entanglement
can be used as an additional resource.

Fig. 1 shows the type of source that we have in mind. A pump pulse coming from below traverses a non-linear crystal
where it can produce correlated pairs of photons into the modes $a_1$ and $b_1$. After the crystal it is reflected
and traverses the crystal a second time, now producing correlated pairs into the spatial modes $a_2$ and $b_2$. The
photon pairs can additionally be entangled in polarization. It is experimentally possible to fix the distance
between the crystal and the mirror such that the phase between the first and the second possibility to create
photon pairs is stable \cite{herzog} and equal to a multiple of $2 \pi$.

\begin{figure}[h!]
\includegraphics[width=3.4in]{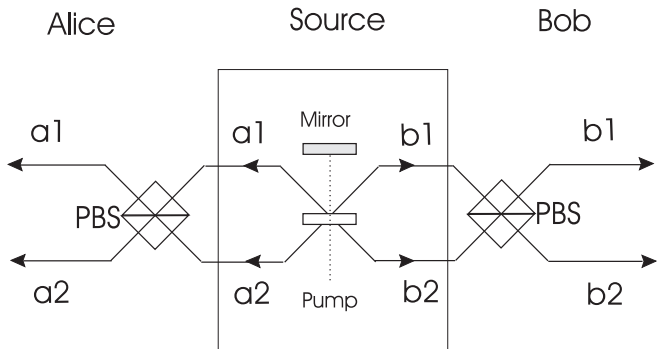}
\caption{Setup for the generation of purified polarization
entanglement with the help of spatial entanglement. The source consists of a non-linear crystal pumped by a laser
pulse. The pump pulse is reflected from a mirror such that it traverses the crystal twice. Photons can be created
in pairs both into the upper modes $a_1$ and $b_1$ and into the lower modes $a_2$ and $b_2$, with a fixed phase
between these two possibilities, leading to spatial entanglement. The photons are additionally entangled in
polarization. In practice, spatial and polarization entanglement will be imperfect when the photons have travelled
from the source to Alice and Bob. In the text we analyze the cases where two and four photons have been produced.
In both cases the setup shown serves to purify the polarization entanglement. On each side the two spatial modes
are combined on polarizing beam splitters. For two photons, one selects those cases where the photons are both in
the upper or both in the lower spatial modes. For four photons, one selects those cases where there is one photon
in each output mode. Then {\it both} the pair of photons in modes $a_1$ and $b_1$ {\it and} the pair in $a_2$ and
$b_2$ have higher polarization entanglement than before.} \label{purpdc}
\end{figure}

Then the situation is approximately described by the Hamiltonian
\begin{eqnarray}
H=\gamma K^+ + \gamma^* K^- =\gamma(a^{\dagger}_{1H} b^{\dagger}_{1H} + a^{\dagger}_{1V} b^{\dagger}_{1V}
 + a^{\dagger}_{2H} b^{\dagger}_{2H} \nonumber\\+ a^{\dagger}_{2V} b^{\dagger}_{2V}) + h.c.,
\end{eqnarray}
where $H$ and $V$ denote vertical and horizontal polarization and we have defined $K^+=a^{\dagger}_{1H}
b^{\dagger}_{1H} + a^{\dagger}_{1V} b^{\dagger}_{1V}
 + a^{\dagger}_{2H} b^{\dagger}_{2H} + a^{\dagger}_{2V} b^{\dagger}_{2V}$. Considering the single-pair state created
 by this source, $K^+|0\rangle$ (we will disregard normalization where it is not essential), one sees that it creates photon pairs that
 are entangled both in polarization and in the spatial modes. In a different notation, the state $K^+|0\rangle$ could be written
 as $({\it 11+22})(VV+HH)$, so there are two qubits on each side, one represented by the polarization modes and one by the
spatial modes, and both the polarization and the spatial qubits are in a singlet state. The total entanglement
content is therefore two ``ebits''.

The four-photon state produced by this source is given by
\begin{eqnarray}
&&(K^+)^2 |0\rangle=(a^{\dagger}_{1H} b^{\dagger}_{1H} +
a^{\dagger}_{1V} b^{\dagger}_{1V}
 + a^{\dagger}_{2H} b^{\dagger}_{2H} + a^{\dagger}_{2V} b^{\dagger}_{2V})^2
 |0\rangle\nonumber\\&&=((a^{\dagger}_{1H} b^{\dagger}_{1H} +
a^{\dagger}_{1V} b^{\dagger}_{1V})^2+(a^{\dagger}_{2H}
b^{\dagger}_{2H} + a^{\dagger}_{2V} b^{\dagger}_{2V})^2
\nonumber\\&&+2(a^{\dagger}_{1H} b^{\dagger}_{1H} +
a^{\dagger}_{1V} b^{\dagger}_{1V}) (a^{\dagger}_{2H}
b^{\dagger}_{2H} + a^{\dagger}_{2V} b^{\dagger}_{2V}))|0\rangle
 \label{4phot}
\end{eqnarray}

One sees that this state contains a component where there is one photon in each spatial mode $a_1$, $a_2$, $b_1$
and $b_2$. But it also has components of comparable magnitude with two photons each in the upper modes $a_1$ and
$b_1$, or two photons each in the lower modes $a_2$ and $b_2$. This shows the quasi-thermal nature of
down-conversion: given two PDC sources, the probability that each emits a pair is of the same order of magnitude as
the probability that one of them emits four photons and the other one doesn't emit any photons at all. As mentioned
above, this is usually perceived as a problem. However, the state (\ref{4phot}) contains a lot of potentially
useful entanglement. Expanding (\ref{4phot}) one easily shows that it is a maximally entangled state in $10 \times
10$ dimensions and thus contains $^2 \log 10=3.32$ ebits, i.e. significantly more than two separate
polarization-entangled pairs.

So far we have been talking about the ideal case of perfect
polarization and spatial entanglement. In practice neither of them
will be perfect. For example, the photons traveling from the
source to Alice's station may suffer depolarization, consisting of
both bit-flip and phase errors, which reduces the polarization
entanglement. The spatial entanglement is affected if the phase
between the two possibilities for creating photons is not exactly
stable. However, the probability for bit-flip errors in the
spatial modes is extremely low, cross-talk between the two spatial
modes on each side can be easily avoided e.g. by having two
separate optical fibers.

Fig. 1 shows the basic setup for our purification scheme. The two spatial modes on each side are combined on
polarizing beam splitters (PBS). This resembles the setup in \cite{pan}, but with a different source. A PBS
transmits horizontally polarized photons and reflects vertically polarized ones. In the language of modes this
corresponds to the transformations $a_{1H} \rightarrow a_{2H}, a_{1V} \rightarrow a_{1V}, a_{2H} \rightarrow
a_{1H}, a_{2V} \rightarrow a_{2V}$, and analogously for the modes on Bob's side. Here we have denoted the spatial
modes behind the PBS by the same names as the original spatial modes.

The basic reason why the setup of fig. 1 performs entanglement purification is the following. On the one hand, the
PBS ensure that photons of different polarization that are originally in the same spatial mode end up in different
spatial modes. On the other hand, photons are always created into corresponding pairs of spatial modes. As a
consequence, selecting certain distributions of photons over the spatial modes allows one to get rid of the cases
where a bit-flip error has occurred in polarization, cf. \cite{bouwmeester}. Phase errors can be purified in a
second step, by first transforming them into bit-flip errors \cite{bennett}. This leads to universal purification
protocols.

Let us first illustrate the purification effect of fig. 1 for the simplest case, where only a single photon pair
has been produced by the source. In the ideal case the state is therefore given by $K^+ |0\rangle=(a^{\dagger}_{1H}
b^{\dagger}_{1H} + a^{\dagger}_{1V} b^{\dagger}_{1V}
 + a^{\dagger}_{2H} b^{\dagger}_{2H} + a^{\dagger}_{2V} b^{\dagger}_{2V}) |0\rangle$. One sees that this state is not changed by
the action of the two PBS, so also after the PBS the photons will be either in the upper modes $a_1$ and $b_1$, or
in the lower modes $a_2$ and $b_2$. But suppose that a bit-flip error in polarization has occurred on the way to
Alice's station, e.g. exchanging $a_{1V}$ and $a_{1H}$. Then after the PBS one of the photons will be in an upper
mode and the other one in a lower mode. Therefore by selecting only those events where both photons are up (one in
$a_1$ and one in $b_1$) or both are down (one in $a_2$ and one in $b_2$), one can purify away all bit-flip errors.

Several remarks are in order. First it is worth noting that for
the case of a single photon pair the above setup is actually a
realization of the purification scheme proposed in \cite{bennett},
which uses CNOT operations on each side. The PBS is an
implementation of the CNOT operation between a spatial-mode and a
polarization qubit, since the spatial mode is flipped or not
flipped as a function of the polarization. This implies that the
above scheme also works if the original spatial entanglement is
not perfect. The more efficient scheme of \cite{deutsch} can also
be realized in this way.

Second, the PBS transform spatial entanglement into polarization
entanglement. To see this, consider the amplitude for finding the
two photons in modes $a_1$ and $b_1$ after the PBS. There are two
ways of reaching this final state. Either the photons can have
come from the two upper modes, then they must have been reflected
by the PBS and thus must both be vertically polarized, or they
came from the lower modes, then they must have been transmitted
and thus be horizontally polarized. If there is a fixed phase
between these two possibilities, i.e. if there was original
spatial entanglement, then one has a polarization-entangled state.

Third, we have stated that a purified polarization-entangled pair is produced in the pairs of modes $a_1-b_1$ or
$a_2-b_2$. With present technology, these good cases can only be selected a posteriori. For purification schemes
involving several steps this means that one will sometimes run the second step although the first step did not
actually produce a pair. Avoiding this kind of inefficiency would require a method for non-destructive detection of
photons.

The above method for the purification of single photon pairs is interesting in its own right because of its great
simplicity. The main experimental requirements are phase stability of the setup and good overlap of the photon
wavepackets on the two PBS. Without good overlap, the polarization of the photons behind the PBS could be inferred
from their temporal characteristics, which means that the polarization entanglement would be affected
\cite{zukowski}.

Let us now turn to the case where four photons are produced by the
source in fig. 1, subsequently referred to as ``four-photon
case''. Ideally one would have the state (\ref{4phot}). As before,
this state is unchanged by the action of the two PBS. The protocol
proceeds by selecting those cases where there is one photon in
each spatial mode behind the PBS, subsequently called the
``four-mode cases''. For the ideal state, this projects onto
\begin{equation}
(a^{\dagger}_{1H} b^{\dagger}_{1H} + a^{\dagger}_{1V} b^{\dagger}_{1V}) (a^{\dagger}_{2H} b^{\dagger}_{2H} +
a^{\dagger}_{2V} b^{\dagger}_{2V}) |0\rangle,
\label{2pairs}
\end{equation}
a state of two independent polarization-entangled pairs, one in
the upper and one in the lower modes. Note that the other terms in
(\ref{4phot}) have all four photons in the upper modes or all four
photons in the lower modes, so they do not lead to even threefold
coincidences. Again, to arrive at the state (\ref{2pairs}),
spatial entanglement has been transformed into polarization
entanglement by the two PBS. Consider the two photons in the upper
modes: they can have been both reflected by the PBS if they are
vertically polarized, or both transmitted if they are horizontally
polarized. Again the polarization entanglement arises from the
fixed phase between these two possibilities.

To understand why the setup has a purifying effect, suppose again that a single bit-flip error occurs in one of the
spatial modes, e.g. mode $a_1$. Then the affected photon is diverted by the PBS from the path that it would take in
the error-less case, and thus there cannot be a photon in each of the four output modes, but there will be only a
three-fold coincidence. Furthermore recall that in the ideal case there are no threefold coincidences, so diverting
a single photon does not turn the cases thrown away in the ideal case into four-fold coincident cases. Therefore by
selecting the four-mode cases one can indeed purify away single bit-flip errors. To study the actual magnitude of
the purification effect, this simple argument has to be supplemented by a detailed calculation, cf. below.

It is important to note that in the present protocol both the
upper and the lower pair of photons can be used. This is in
contrast to the scheme of \cite{pan}, where only one of the output
pairs was useful. The reason for this substantial improvement in
entanglement output is that we have succeeded in using spatial
entanglement as an additional resource. In contrast to usual
purification schemes, where entanglement is concentrated into a
single pair, while the other pair is discarded, here the spatial
entanglement is concentrated into polarization entanglement, and
in the four-mode case all photons are kept. However, photons are
indeed discarded, since a four-mode event occurs only in about 40
percent of the four-photon cases (the exact value depending on the
exact four-photon state).

It is interesting to compare the present protocol to the performance of the PBS scheme for single-pair sources
without spatial entanglement \cite{pan}, i.e. for an initial state approximately corresponding to (\ref{2pairs}),
but with imperfect polarization entanglement. Selecting four-mode cases behind the PBS, which happen with a
probability of 50 percent, projects (\ref{2pairs}) onto
\begin{equation}
(a^{\dagger}_{1H}  a^{\dagger}_{2H}b^{\dagger}_{1H} b^{\dagger}_{2H} + a^{\dagger}_{1V} a^{\dagger}_{2V}
b^{\dagger}_{1V}
 b^{\dagger}_{2V}) |0\rangle,
\label{ghz}
\end{equation}
which has 1 ebit of entanglement between Alice and Bob. This is an upper bound for the real case where the original
polarization entanglement is less than perfect. Thus for true single-pair sources the protocol outputs one purified
entangled pair in 50 percent of the four-photon cases, where the entanglement fidelities before and after
purification are related in the same way as for the protocol of \cite{bennett}, cf. our discussion in \cite{pan}.
The well-known S-shaped curve describing this relationship is plotted in figure 2. Here we have defined the
entanglement fidelity $F$ of a mixed state $\rho$ as $F=\langle \psi |\rho|\psi\rangle$, where
$|\psi\rangle=\frac{1}{\sqrt{2}}(|H\rangle|H\rangle+|V\rangle|V\rangle)$ is the desired maximally entangled pure
state.

On the other hand, for our present scheme that uses spatial entanglement the probability of a four-mode case behind
the PBS is approximately 0.4, and every four-mode case corresponds to two purified pairs, where the relationship
between initial and final polarization entanglement fidelities is also plotted in figure 2. Below we will describe
in more detail how these curves were obtained. One sees that in the ideal case the new curve is always above the
curve of \cite{bennett}, while for reasonably good spatial entanglement the new curves are still substantially
above the ideal curve of \cite{bennett} over a wide range of initial fidelities. Let us stress again that moreover
there are {\it two} output pairs instead of one. It is particularly remarkable that there is no lower threshold for
purification, in contrast to the previous schemes of \cite{bennett} and \cite{pan}, which had fidelity thresholds
of $F=1/2$, as can be seen in figure 2. Also this feature is understandable because of the presence of spatial
entanglement, which is converted into polarization entanglement by the PBS.

The above comparison to the case of single-pair sources shows clearly that for the present protocol the thermal
character of the down-conversion source is not destructive, but actually helpful, provided that there is a stable
phase between the two possible photon creation events. Parts of the final four-mode four-photon amplitude come from
cases where two photons were created into the same spatial mode on both sides. These contributions account for the
difference between (\ref{2pairs}) and (\ref{ghz}), and thus for the additional unit of entanglement (in the ideal
case).

\begin{figure}
\includegraphics[width=0.8 \columnwidth]{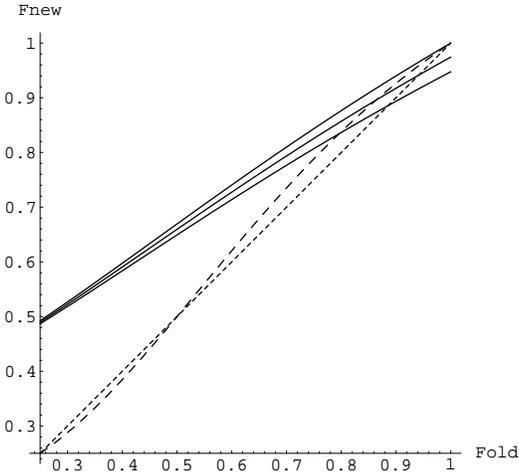}

\caption{This plot shows the polarization entanglement fidelity
after purification as a function of the fidelity before purification for different schemes and parameter values.
The three full-line curves show the performance of our scheme, where the top one is for perfect spatial
entanglement, i.e. $r=\cos \phi=1$, and the two others are for $r=\cos \phi=0.95$ and $r=\cos \phi=0.9$. The dashed
S-shaped curve shows the performance of the protocols of refs. 1 and 4 under ideal conditions. The straight line
corresponds to equal input and output fidelities. It is important to note that the new scheme produces two
entangled pairs instead of one. One sees that the new protocol does not have a lower threshold fidelity, while the
upper threshold is determined by the quality of the spatial entanglement.}
\label{fids}
\end{figure}

Let us now explain how exactly the curves in figure 2 that refer to our new protocol are obtained. Neither spatial
nor polarization entanglement are supposed to be perfect. Instead we assume that our source can be characterized by
a Hamiltonian
\begin{equation}
H=\gamma ((a^{\dagger}_{1H} b^{\dagger}_{1H} + a^{\dagger}_{1V} b^{\dagger}_{1V})+r e^{i\phi} ( a^{\dagger}_{2H}
b^{\dagger}_{2H} + a^{\dagger}_{2V} b^{\dagger}_{2V}) )+ h.c.,
\label{realsource}
\end{equation}
where the relative probability of producing a pair into the lower as opposed to the upper spatial modes is
determined by the coefficient $r$, and the phase between these two possibilities is denoted by $\phi$. Furthermore
we assume that on their way to Alice's station the photons are subjected to a partially depolarizing channel.

The fully depolarizing channel on a subsystem $A$ of a composite
system $AB$ is defined as $C_o: \rho_{AB}\rightarrow
\frac{1}{d}\openone_A \otimes \mbox{Tr}_A \rho_{AB}$, where $d$ is
the dimension of system $A$. In component notation this can be
written as $|i\rangle \langle j|\rightarrow \delta_{ij}
\frac{1}{d} \sum \limits_{n=1}^d |n\rangle \langle n|$.
In our calculation we have assumed that the channel affects polarization but not photon number, so that for example
the fully depolarizing channel in a given spatial mode, e.g. $a_1$, affects the density matrix elements  in the
following way:
\begin{eqnarray}
&&|0\rangle \langle 0| \rightarrow |0\rangle \langle 0| \nonumber\\ &&|1_H,0_V\rangle \langle 1_H,0_V|\rightarrow
\frac{1}{2}(|1_H,0_V\rangle \langle 1_H,0_V|+|0_H,1_V\rangle \langle 0_H,1_V|)\nonumber\\
&&|2_H,0_V\rangle \langle 2_H,0_V| \rightarrow \frac{1}{3}(|2_H,0_V\rangle \langle
2_H,0_V|\nonumber\\&&+|1_H,1_V\rangle
\langle 1_H,1_V|+|0_H,2_V\rangle \langle 0_H,2_V|),
\end{eqnarray}
and analogously for all other diagonal matrix elements, while all off-diagonal elements are transformed into zero.
Here we have defined $|1_H,0_V\rangle=a^{\dagger}_{1H}|0\rangle$ etc.

We define the partially depolarizing channel $C_s$ as the application of the fully depolarizing channel with
probability $1-s$, while the system remains undisturbed with probability $s$. In our calculations we consider
states created by the source (\ref{realsource}), which is characterized by $r$ and $\phi$, and then apply
depolarizing channels $C_s$ in the spatial modes $a_1$ and $a_2$, which could e.g. correspond to a situation where
the distance from the source to Alice is much larger than the distance from the source to Bob. The polarization
entanglement fidelity of individual photon pairs created in this way is $\frac{(1+3s)}{4}$, which is the fidelity
before purification defining the x-axis in fig. 2. Fig. 2 shows that effective purification can be achieved for
realistic values of $r$ and $\phi$. The final fidelity that can be achieved is determined by the quality of the
spatial entanglement.

A first experimental realization of the present scheme is under way. The scheme is scalable in principle. Several
sources of the type of fig. 1 can produce photons in parallel, which  can then be fed into stacked arrays of
polarizing beam splitters. Phase stability of the whole setup has to be achieved. The methods of the present work
can be adapted to the case of energy-time entanglement \cite{brendel}, which allows to go to longer distances. True
long-distance quantum communication protocols will probably also require the capability of storing photons in order
to overcome the problem of photon loss. A protocol combining photons and atomic ensembles has recently been
proposed \cite{duan}.

We are grateful to A. Zeilinger for triggering our interest in
purification. We thank T. Tyc for useful discussions. This work
was supported by the QIPC program of the European Union and by the
Austrian Science Foundation (FWF).

\end{document}